\documentclass[a4paper,11pt]{article}

\pdfoutput=1 
\usepackage[T1]{fontenc} 

\usepackage[]{caption}
\usepackage{color,xcolor,graphicx,amsmath, amssymb,mathtools
, physics,ulem}

\usepackage{upgreek}			
\usepackage{bm}					
\usepackage{booktabs} 

\newcommand{\be}{\begin{eqnarray}}
\newcommand{\ee}{\end{eqnarray}}

\newcommand{\beq}{\begin{equation}}
\newcommand{\eeq}{\end{equation}}

\definecolor{shilamagenta}{rgb}{0.8, 0.0, 0.8}

\definecolor{shilagreen}{rgb}{0.0, 0.5, 0.0}
\definecolor{shilacyan}{rgb}{0.0, 0.58, 0.71}

\definecolor{midnightblue}{rgb}{0.1, 0.1, 0.44}

\usepackage{hyperref}

\hypersetup{citecolor=blue,        
    linkcolor=red,  
    filecolor=magenta,      
    urlcolor=shilacyan  
}

\usepackage{cleveref}
\usepackage{slashed}

\begin{document}

\title{Dark Matter from Evaporating Primordial Black Holes in the Early Universe}

\author{Pratik Chattopadhyay
$^{a}$\footnote{{\bf e-mail}: \href{mailto:pratikpc@gmail.com}{pratikpc@gmail.com} },
Arnab Chaudhuri$^{b,c}$ \footnote{{\bf e-mail}: \href{mailto:arnabchaudhuri.7@gmail.com}{arnabchaudhuri.7@gmail.com} }
and
Maxim Yu. Khlopov$^{d}$\footnote{{\bf e-mail}: \href{mailto:khlopov@apc.in2p3.fr}{khlopov@apc.in2p3.fr}}
\\
$^a$ \small{School of Mathematical Sciences,} 
\small{ University of Nottingham.}\\
$^b$ \small{Discipline of Physics,} 
\small{ Indian Institute of Technology, Gandhinagar.}\\
$^c$ \small{Division of Science,} 
\small{ National Astronomical Observatory of Japan, Mitaka, Tokyo.}\\
$^d$ \small{Research Institute of Physics,} 
\small{Southern Federal University}\\
\small{Virtual Institute of Astroparticle Physics}\\
\small{National Research Nuclear University “MEPHI”}
}

\date{}
\maketitle

\begin{abstract}
Primordial Black Holes (PBH) could dominate in the early universe and, evaporating before Big bang Nucleosynthesis, can provide new freeze in mechanism of dark matter (DM) production. The proposed scenario is considered for two possible mechanisms of PBH formation and the corresponding continuous PBH mass spectra so that the effect of non-single PBH mass spectrum is taken into account in the results of PBH evaporation, by which PBH dominance in the early universe ends. We specify the conditions under which the proposed scenario can explain production of dark matter in very early Universe.
\end{abstract}

\section{Introduction}
\label{sec:intro}

The Standard Model (SM) of particle physics has been incredibly successful in describing the behavior of elementary particles and their interactions, except for those involving gravity. However, there are still some issues with the SM, including its inability to explain the particulate nature of Dark Matter (DM). Dark Matter is a type of matter that does not interact with light or any other form of electromagnetic radiation, which makes it invisible to telescopes and other forms of traditional observation \cite{Planck:2018vyg, WMAP:2012nax, XENON:2018voc, Dark Matter, Mohan:2022kvb}. Despite this, the presence of DM is evident in the universe, as it accounts for more than one fourth of the energy budget.

There have been proposals for several mechanisms for the production of DM, including the WIMP (Weakly Interacting Massive Particle) and FIMP (Feebly Interacting Massive Particle) scenarios, which require DM candidates to have a coupling with the SM sector. However, it is also possible for DM particles to be produced from the evaporation of Primordial Black Holes (PBHs) via Hawking radiation, which would make them completely isolated from the SM sector. This means that they would be immune to various direct detection and collider constraints that apply to other forms of DM \cite{XENON:2018voc,LUX:2016ggv,PandaX-II:2017hlx, Boveia:2018yeb}.

As is well know since a long time now, black holes are objects that are formed due to gravitational collapse of stars. There are different kinds of black hole solutions in Einstein gravity and some of them are Schwarzschild, Kerr and Reissner-Nortsdom. For a brief review, see \cite{BH1, BH2}. Primordial black holes (PBH) are black holes which formed in the early universe due to accumulation of over-dense regions as a result of quantum fluctuations or strong nonhomogeneity of very early Universe, reflecting various aspects of physics beyond the Standard Model of fundamental interactions (see e.g. \cite{PBHrev} for review and references). There are various scenarios for the formation of PBHs and some of them are: collapse from inhomogeneities \cite{Kokubu}, collapse of cosmic strings \cite{Jenkins}, bubble collisions \cite{Jung}, from tiny bump in the inflaton potential \cite{Mishra} and collapse of domain walls \cite{Liu}. 
\\
PBHs can be produced in the early universe from quantum fluctuations and later on dominate the energy density of the universe, only to eventually evaporate away completely by radiating DM particles along with other SM states before the advent of Big Bang Nucleosynthesis (BBN). 

Apart from the standard cosmology, a lot of modern and relatively new cosmological models have surfaced recently. One of the most successful models is the model of bouncing cosmology. The Big Bounce hypothesis is a cosmological model for the origin of the known universe. It was originally suggested as a phase of the cyclic model or oscillatory universe interpretation of the Big Bang, where the first cosmological event was the result of the collapse of a previous universe. A more detailed explanation of the modern universe comes in the form of Bianchi models which take into the anisoptropies of the universe. There are 9 Bianchi models. \cite{cosmo1, cosmo2, cosmo3, cosmo4, cosmo5}

The production of DM from PBH evaporation in the context of a single PBH has been widely studied in the literature. However, a recent study \cite{Cheek:2022mmy} has explored the possibility of DM production from PBH mass and spin distributions, where the PBHs are produced simultaneously at the moment when the PBH corresponding to the peak mass value is produced.

In a typical scenario of DM production from a single PBH evaporation, a DM particle in the rather wide mass range ($1-10^9$) GeV cannot satisfy the relic density ($\Omega h^2=0.120\pm0.001$ as per Planck data \cite{Planck:2018vyg}) in the PBH dominated region of parameter space \cite{Cheek:2021cfe} due to BBN constraints. However, it is possible that this constraint can be lifted in the presence of multiple PBHs. In this work, we investigate the similar phenomenon with a non-monochromatic PBH distribution where the PBHs are formed corresponding to the peak value of the PBH. We compute the ratio of the relic abundance to that of the relativistic particles corresponding to two different PBH spectrums. The results come out to be the same as the ratio just depends on the peak value of the distribution. However, if we allow for distributions with different peaks, the results vary. \\~\\This paper is arranged as follows. In the next section, we describe two different mass spectrums for PBH, both of which are extended ones. One of them is motivated from phase transitions at the inflationary stage and the other one arises due to collapse of domain walls. In section 3, we describe the formation of dark matter and present some estimations which gives us the ratio of DM particle densities to that of relativistic particle densities. In section 4, we conclude with a discussion of the results.


\section{Primordial Black Holes and Mass Spectrum}
In this work, we focus on short-lived PBHs that evaporated early in the universe's expansion, prior to the onset of BBN. These short-lived PBHs have significant implications in the present universe. Firstly, their decays can introduce a substantial influx of entropy into the plasma, potentially diluting any preexisting asymmetry. Secondly, the decay of PBHs can generate a baryon asymmetry. Lastly, PBH evaporation can result in the production of dark matter. Specifically, we investigate the production of dark matter in this study and demonstrate that, by appropriately selecting parameters, relics generated from PBH evaporation can make a substantial contribution to the dark matter density.

Initially, we assume the universe is in a radiation-dominated (RD) stage, where most of the cosmological matter consists of relativistic particles. During this epoch, the energy density is given by:
\be
\rho_{rel} ^{(1)}= \frac{ 3 m_{Pl}^2}{32 \pi t^2} .
\label{rho-rel}
\ee
where $m_{Pl}$ is the Planck mass. The cosmological scale factor evolves with time according to:
\be 
a_{rel} (t) = a_{in} \, \left(\frac{t}{t_{in}}\right)^{1/2}.
\label{a-rel}
\ee
If the number density of PBHs is sufficiently high and their mass allows them to persist until they begin dominating the universe, the energy density after a certain time $t_1$ is given by:
\be
\rho_{nr} = \frac{ m_{Pl}^2}{6 \pi (t+ t_1) ^2}.
\label{rho-nr}
\ee
where $t_1$ is determined from the condition of equality between the $\rho_{rel}$ and $\rho_{nr}$ at the equilibrium time $t_{eq}$. The value of $t_1$ can be calculated as $t_1 = t_eq / 3$. The equilibrium time $t_{eq}$ can be obtained from the equations:
\be
a_{in}/ a_{eq} = (t_{in} / t_{eq})^{1/2} = \rho_{BH}^{in} / \rho_{rel}^{in}.
\ee
Following a certain time $t_2$, the PBHs undergo evaporation and produce relativistic matter, causing the expansion regime to return to the relativistic stage when all or a significant portion of the PBHs have evaporated. The energy density during this stage is given by:
\be
\rho_{rel}^{(2)} = \frac{ 3 m_{Pl}^2}{32 \pi( t+t_2)^2}.
\label{rho-rel2}
\ee 
where $t_2$ is determined by the condition of equality between $\rho_{nr}$ (\ref{rho-nr})
and $\rho_{rel}^{(2)}$ (\ref{rho-rel2}) at the time of PBH decay, $t=\tau_{BH}$.

\subsection{Parameterization of the extended mass spectrum}
In this study, we investigate an extended mass spectrum for PBHs. The extended mass spectrum is generally described by the equation:
\be 
\frac{dN_{BH}}{dM} = f(M,t) ,
\label{dN-dM}
\ee
where $N_{BH}$ represents the number density of PBHs, $M$ denotes the mass, and $t$ corresponds to the evolving time. Since these PBHs exhibit non-relativistic behavior, their differential energy density with respect to mass can be expressed as:
\be 
\frac{d\rho_{BH}}{dM} \equiv \sigma(M,t) = M f(M,t).
\label{rho-BH-2}
\ee
We make the assumption that although the PBHs are formed through conventional mechanisms, they possess a broader spectrum, as suggested in earlier works ~\cite{DKK,INN}.

In our scenario, we consider a range of PBHs with number and energy densities confined between $M_{min}$ and $M_{max}$. The minimum PBH mass, $M_{min}$, must be greater than a lower bound that ensures the assumption $\tau_{BH} \geq t_{eq}$ holds. The maximum PBH mass, $M_{max}$, is determined by the requirement that PBH evaporation does not distort the well-established results of BBN-theory.

To parameterize the PBH mass, we introduce a dimensionless parameter, $x$, such that
\be \label{pam-x}
M= x {M}_{0},
\ee
with $M_0$ representing the mean value of the distribution, we define it as the value at which $\sigma (M,t)$ and $x$ become nonzero within the limits:
\be
x_{min} \equiv M_{min}/M_0 \le x \le x_{max} \equiv M_{max}/M_0 .
\label{x}
\ee

We introduce the dimensionless "time" parameter $\eta$, defined as $\eta = t / \tau(M_0)$, where $\tau(M_0)$ represents the lifetime of a PBH with mass $M_0$. Here, $\tau_0$ denotes the lifetime of a PBH with the mean mass $M_0$. Due to the varying lifetimes of individual PBHs, their masses and creation times also differ. The evolution of the differential energy is then described by:
\be 
\dot \sigma (M, t)=-\left[ 3H+\Gamma(M)\right] \sigma (M,t).
\label{en-den}
\ee 
The quantity $\Gamma(M)$ is defined as $\Gamma(M) = 1/\tau(M)$, where $\tau(M)$ is given by $\tau(M) = (3 \times 10^3 N_{eff}^{-1} M_{BH}^3 m_{pl}^{-4})$. Here, $C$ is approximately 30 and $N_{eff}$ represents the effective number of particle species with masses smaller than the black hole temperature. Additional information can be found in the reference \cite{DKK}. Thus, the expression for $\Gamma(M)$ can be written as:
\be
\Gamma(M)=\frac{m_{Pl}^4}{(C M^3)}.
\ee
In the units of $\eta$, (\ref{en-den}) can be written as:
\be 
\frac{d\sigma}{d\eta}  \equiv \sigma' =-\left[3H\tau_0+\left(\frac{M_0}{M}\right)^3\right] \sigma .
\label{en-den1}
\ee
The initial value of $\eta$ is the moment of PBH formation, depends on $M$ and has the form:
\be 
\eta_{form} (M) =\frac{m_{Pl}^2 M}{C{M_0}^3} .
\label{eta-in-M}
\ee
Evidently $\sigma (M) = 0$ when $\eta (M) <\eta_{form}$. 

The energy evolution of the relativistic matter can be written as:
\be
\frac{d \rho_{rel}}{d\eta} \equiv \rho_{rel}'  
= - 4H\tau_0  \rho_{rel}  + \int dM   (M_0/M)^3 \sigma (M) .
\label{rho-rel-eq-2}
\ee
The red-shift factor as a function of $\eta$ normalised to the value of the scale factor at the moment of
the least massive PBH formation: 
\be
z (\eta) = a(\eta) / a\left[ \eta_{form}(M_{min}) \right] .
\label{z-2}
\ee
The temporal evolution of $z$ is governed by the Hubble parameter and can be written as:
\be
\frac{dz}{d\eta} = H \tau_0 z
\label{dz-deta-1}
\ee
with the Hubble parameter following:
\be 
\frac{3H^2  m_{Pl}^2}{8\pi} = \rho_{rel} + \rho_{BH} =
\rho_{rel}+ \int dM \sigma (M),
\label{hubble-of-M}
\ee
Eqn. (\ref{en-den1}) follows the following solution:
\be
\sigma (M,\eta) = \theta \left(\eta - \eta_f \right) \sigma (M,\eta_{f}) 
\exp \left[ \left( \eta_f - \eta \right)
\left(\frac{M_0}{M}\right)^3 \right] \left( \frac{z(\eta_{f})}{ z(\eta)} \right)^3 ,
\label{sol-r}
\ee
where for simplicity we chose $\eta_f \equiv \eta_{form} (M)$. The theta function ensures that the function vanishes outside the bounds. 

The initial value of the PBH energy density at the moment of creation depends on the factor $\epsilon (M)$ which is defined as:
\be
\epsilon (M)=\frac{\rho_{PBH}^{in}}{\rho_{rel}^{in}}.
\ee
And hence the initial value of the PBH energy density takes the form:
\be
\sigma (M, \eta_f(M) )= \epsilon (M) \rho_{rel} (\eta_{f} (M) ) / M.
\label{r-in-of-M}
\ee
The term $\epsilon(M)$ is influenced by the specific scenario of PBH formation and becomes negligible outside the bounds of the PBH mass spectrum. We make the assumption that within the interval $\eta_f (M_{min}) < \eta < \eta_f (M_{max})$, the total fraction of PBH mass density is significantly smaller compared to the energy density of relativistic matter. As a result, the expansion regime remains undisturbed and remains in the relativistic stage. The energy density of relativistic particles at the time of the first (lightest) PBH formation can be expressed as:
\be
\rho_{rel} (t_{in}) = \frac{3}{32\pi} \,\frac{m_{Pl}^6}{M_{min}^2} .
\label{rho-rel-in}
\ee 
If the energy density of PBH remains much smaller than that of relativistic matter until the formation of the heaviest PBHs, the last term on the right-hand side of equation (\ref{rho-rel-eq-2}) is disregarded. In this case, within the time interval $\eta (M_{min}) < \eta <\eta(M_{max})$, the energy density $\rho_{rel}$ equals
\be
\rho_{rel} =  \frac{3}{32\pi} \,\frac{m_{Pl}^6}{M_{min}^2}\frac{1}{z(\eta)^4} .
 \label{rho-rel-of-eta}
 \ee
And the differential energy density becomes:
\be
\sigma(M, \eta) = \frac{3 m_{Pl}^6}{32\pi M M^2_{min}}\,\frac{\epsilon (M)}{z(\eta_f(M)) }
\frac{\theta (\eta - \eta_f(M))}{z^3(\eta) \exp\left[ (M_0/ M)^3  (\eta - \eta_f(M) )  \right] } .
\label{sigma-of-eta}
\ee
In this  equation $\eta$ runs in the limits $\eta (M_{min}) < \eta < \eta (M_{max}) $ or 
$\eta_f (M)< \eta < \eta (M_{max})$, depending upon which lower limit is larger. 

Since $ (M_0 / M)^3 \eta_f (M) = m_{Pl}^2 / (C M^2) \ll 1$, for any $\eta$, we may expand the exponent as
\be
 \exp \left[ -(M_0/ M)^3  (\eta - \eta_f(M) )  \right]   =  \exp\left[ - (M_0/ M)^3  \eta   \right]  (1+  m_{Pl}^2 / (C M^2) )
 \label{exp-expand}
 \ee
To simplify the numerical calculations involved in integrating over the variable $M$ and obtaining desired results, we can consider simplified forms of the initial mass distribution of the PBH. This allows us to evaluate the integrals over $M$ analytically, as demonstrated in the following section (\ref{spectrum}).
In the next step, we assume the following form of the function:
\be
F(x)=\epsilon (M)/z(\eta_f(M)).
\label{F-def}
\ee
The function is restricted between $x_{min} = (M_{min}/M_0)$ and $x_{max} = (M_{max}/M_0)$, as stated in equation (\ref{r-in-of-M}). Here, $\epsilon(M)$ represents the fraction of the energy density of PBH with mass $M$ at the time of PBH formation.
To simplify the analysis, we assume that $F(x)$ is a polynomial function of integer powers of $x$, although this assumption is not required.
\subsection{Extended Mass Spectrum-I} \label{spectrum}
Phase transitions at the inflationary stage can lead to spikes in the spectrum of density fluctuations, strongly increasing the probability of PBH formation in some mass interval \cite{Chaudhuri:2020wjo} and \cite{Dolgov:2020xzo} (see e.g. \cite{PBHrev} for review and references). To illustrate this possibility we take by hands 
an interesting form of the spectrum given by 
\be \label{LN}
F(x)=\frac{\epsilon_0}{N}a^2b^2(1/a-1/x)^2(1/x-1/b)^{2}.
\ee 
where we take 
\be \label{para}
a=1, b=50, \epsilon_0=10^{-12}, N=32\times 10^{-12}.
\ee 
This function vanishes at $x =x_{min}=a$ and $x =x_{max}=b$, with vanishing derivatives at these points. The function reaches it's maximum value at $x_0=2ab/(a+b)$.
\begin{figure}[!htbp]
  \centering
  \begin{minipage}[b]{0.45\textwidth}
    \includegraphics[width=\textwidth]{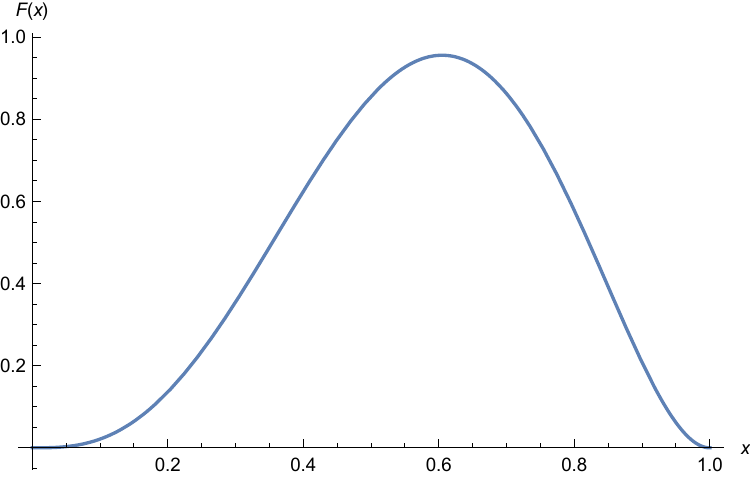}
  \end{minipage}
  \hspace*{.1cm}
  \begin{minipage}[b]{0.45\textwidth}
    \includegraphics[width=\textwidth]{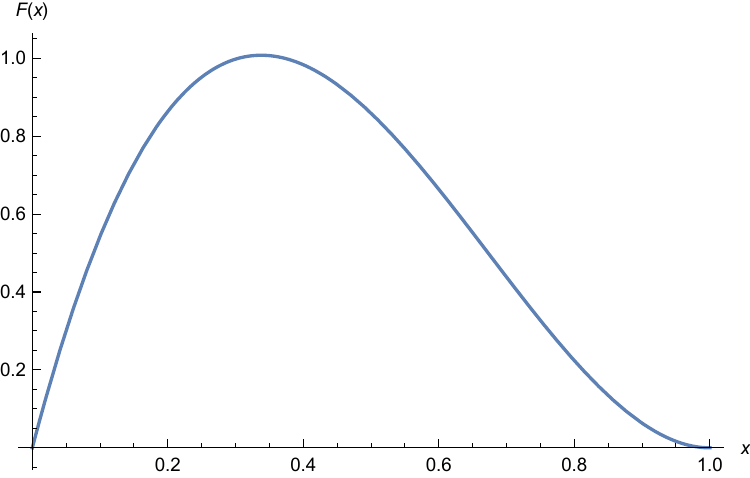}
      \end{minipage}
    \hspace*{.1cm}
  \caption{The extended mass spectrum-I corresponding to Eq. \ref{LN} is shown. The left panel corresponds to the parameters given in \ref{para} while the parameters for the right panel is $N=10^{-10}, a=40~\rm{and}~b=10$. $\epsilon_0=10^{-12}$ is fixed for both the plots. As it is expected, the value of $x$ corresponding to the peak value of the spectrum changes with the plot. It is also evident that the first graph is skewed towards the right while second is skewed towards the left.}
  \label{extended-1}
 \end{figure}
\subsection{Extended Mass Spectrum-II}
Let us now consider the second mass spectrum which arises due to PBH formation from the collapse of domain walls. We first give some estimates for the minimal and maximal mass of the PBHs using the parameters of the scalar field model. Let us recall that the width of the domain wall is inversely proportional to the mass of the scalar field $\phi$
\be 
l\sim \frac{f}{\Lambda^2}\sim \frac{1}{m_{\phi}}
\ee 
where $m_{\phi}$ is the mass of the scalar particle.
The PBH formation occurs when the gravitational radius of the fluctuating field is larger than the width of the domain wall, i.e, 
\be
\label{grv}
\tau_g > l
\ee
The above estimate can be interpreted in terms of the mass of the PBH in the following way. The gravitational radius $r_g$ is given as follows 
\be 
r_g = \frac{2M}{m_{pl}^2}
\ee 
Putting the above value in (\ref{grv}), we get the following inequality 
\be 
\frac{2M}{m_{pl}^2} > \frac{f}{\Lambda^2}
\ee 
This inequality in turn gives the estimate of the minimal mass of the PBH in terms of the parameters $f$ and $\Lambda$. The minimal mass is given by 
\be 
M_{min} = f\Big(\frac{m_{pl}}{\Lambda}\Big)^2
\ee 
To satisfy the BBN constraint, the maximum mass of the PBHs should be more than $10^6g$ . Thus, we need to choose the parameters judiciously such that \cite{Carr:2020gox} and \cite{PBHrev}
\be 
M_{min}=f\Big(\frac{m_{pl}}{\Lambda}\Big)^2 > 10^6 \rm{g}.
\ee 
The above condition gives us the estimate that some of the black holes should evaporate before the BBN. However, if we want to consider the case that all such PBHs evaporate, we should set a condition on the maximal mass of the PBHs. The maximal mass of the PBHs is estimated by the condition that the wall starts to dominate when it enters the horizon. We better avoid this dominance because of the following reason. Initially there are some pieces of the universe in which there are walls and matter/radiation. If walls start to dominate in the sense that the energy density of the walls is greater than that of the matter/radiation, it corresponds to superluminal expansion. The maximal mass obtained from this constraint is given by 
\be 
M_{max}=M_{min}\Big(\frac{m_{pl}}{f}\Big)^2
\ee 
We consider the common scenario for both of these spectra where the entire spectrum of PBHs completely evaporate prior to BBN. This is given by the condition that 
\be 
M_{max}=M_{min}\Big(\frac{m_{pl}}{f}\Big)^2<10^9g
\ee 
It is not necessary that the maximal mass of the PBHs follow the above condition. Indeed, we can have only a part of the whole spectrum to evaporate that could give rise to DM particles. However, there are many sensitive probes based on the estimation of light elements abundance which compels us to put an upper bound to the maximal mass of PBHs as is given above. To this end, let us give some quick estimates. In order to have a consistent inflationary regime, we take the parameter $f$ and $\Lambda$ to be \be 
f = 10^{14} \textrm{GeV} ~~\rm{and}~~\Lambda=10^{10}\rm{GeV}.
\ee  
These values simply imply that it does not lead to over production of gravitino and other particles of that sort and confine ourselves only to stable relics which can be primarily dark matter particles. As can be seen below, the number density of the PBH can be pretty high and it depends on the peak and the character of the spectrum, decreasing with the increase of PBH mass. Unlike the PBHs which are formed by ZN mechanism, which automatically supports a spherical symmetry, the PBHs formed from domain wall collapse do not have any symmetry and hence do not have any particular analytic form . (\ref{spect2}) is the result of an extrapolation of the Figure $2$.
In fact the approximate analytical form the spectrum can expressed as:
\be \label{spect2}
F(x)=1.19628 - 0.96 x - 0.13 x^2 + 0.18 x^3 - 0.022 x^4 - 
 0.13 x^5 + 0.078 x^6 
\ee
The spectrum in consideration here is shown in the following figure 2. Details about this spectrum can be found in \cite{mksrs}.
\begin{figure}[htp]
    \centering
\includegraphics[width=8cm]{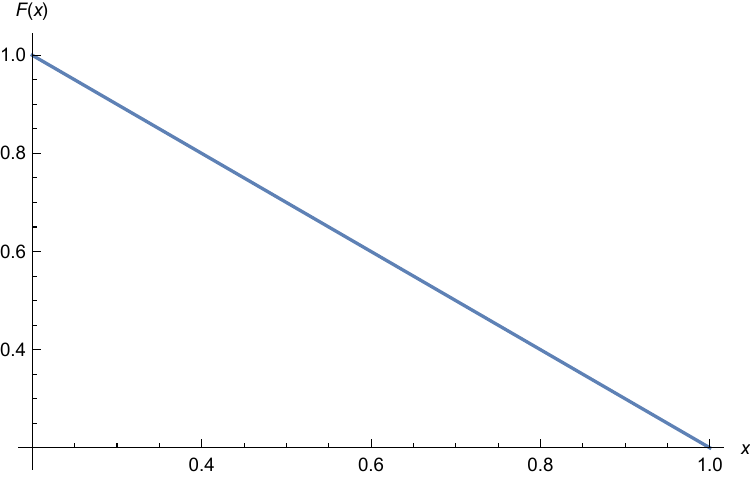}
\label{DWC-1}
\caption{The extended mass spectrum as a result of PBH formation from domain wall collapse.}
\end{figure}

\section{Dark Matter Formation}
Let us now estimate the density of stable dark matter relics produced in PBH evaporation and show their contribution to the cosmological dark matter is significant in some cases whereas insignificant in others. To this end, we present some simple estimates and correspondingly some numerical values. Note, we work only with the Extended Mass Spectrum-I here. At the end of the calculation, we present two tables, one for each spectrum which gives the ratio of the DM particle densities to that of relativistic particle densities. \\~\\
Consider the Extended Mass Spectrum-I. We parameterize the mass of PBH using the dimensionless parameter $x$. Thus we have 
\be 
x=\frac{M}{M_0}
\ee 
where $M$ is the PBH mass and $M_0$ is the peak value of the mass density distribution. At the peak value of $F(x)$, we have $x=0.6$. Let us take $M_0=10^6$g as the first case. The moment of PBH production with mass $M$ is:
\be 
t_{in}=\frac{M}{m_{pl}^2}=\frac{M_6\times 10^6}{(2.18\times 10^{-5})^2}
\nonumber\\
=5.3\times 10^{-33}M_6 \rm{sec}.
\ee 
where $M_6=M (g)/10^6 (g)$
\\~\\
We assume that PBHs make a small fraction of the energy density of relativistic matter at the moment of production. Thus, the energy density and number density at $t=t_{in}$ are 
\be 
\rho^{(in)}_{BH}=\frac{3\epsilon}{32\pi}\frac{m_{pl}^6}{M^2},~~n_{BH}^{(in)}=\frac{3\epsilon}{32\pi}\frac{m_{pl}^6}{M^3}
\ee 
where $\epsilon<<1$.
The energy density of relativistic matter at $t=t_{in}$ is 
\be 
\rho^{(in)}_{rel}=\frac{3}{32\pi}\frac{m_{pl}^6}{M^2}=\frac{\pi^2g*^{(in)}}{30}T^4_{(in)},
\ee 
where $g*^{(in)}\sim 100$ is the number of relativistic species at $T=T_{in}\sim 1.72\times 10^{12}$ GeV/$\sqrt{M_6}$. The ratio of PBH number density to that of relativistic particles at the moment of creation is estimated to be 
\be 
r_{in}=\frac{n^{(in)}_{BH}}{n^{(in)}_{rel}}=\frac{\rho^{(in)}_{BH}}{\rho^{(in)}_{rel}}\frac{T_{in}}{0.3M}=10^{12}\epsilon\times M_6^{-3/2}=5.73\epsilon_{12}M_6^{-3/2}
\ee 
The average distance between PBHs at the moment of their creation is 
\be 
d_{in}^{BH}=(n^{(in)}_{BH})^{-1/3}=11.62\times 10^{-17}M_6\epsilon_{12}^{-1/3} \rm{cm}.
\ee 
Here $\epsilon_{12}=10^{12}\epsilon$. At the moment of equilibrium, the distance of BH separation was 
\be 
d_{eq}^{BH}=d_{in}^{BH}/\epsilon_{12}=11.62\times 10^{-5}M_6\epsilon^{-4/3}
\ee 
The temperature of relativistic matter at equilibrium moment was 
\be 
T_{eq}=\epsilon T_{in}S_{eq}^{1/3}=3.67\epsilon_{12}M_6^{-1/2} \rm{GeV},
\ee 
where $S_{eq}$ is the ratio of the number of particle species at $T=T_{in}$ to that at $T=T_{eq}$, which is $\approx 10$.\\~\\
The universe expanded in a relativistic regime before equilibrium was attained and the scale factor in such a universe rose as $a(t)\sim t^{1/2}$. The equilibrium is reached at the moment of time 
\be 
t_{eq}=t_{in}/\epsilon^2=5.3\times 10^{-33}M_6\epsilon^{-2}=5.3\times 10^{-9}\epsilon_{12}^{-2}M_{6} sec
\ee 
After this time, till the moment of decay,
\be 
t=\tau=30 M_{BH}^3/m_{pl}^4=30 M_6^3\times 10^{18}\times\frac{1}{(2.18)^4\times10^{20}}\times \frac{1}{8.53\times 10^{47}} \rm{sec}\nonumber\\
=1.5\times 10^{-10}M_6^3 \rm{sec}
\ee 
the universe expanded in matter dominated regime with the scale factor going like $a(t)\sim t^{2/3}$. SO during this stage, the scale factor rose as 
\be 
z(\tau)=\Big(\frac{\tau}{t_{eq}}\Big)^{2/3}=0.02(\epsilon_{12}\times M_6)^{4/3}
\ee  
Correspondingly, the energy density of PBHs just before their moment of decay is larger than the energy density of the relativistic background and the amount is calculated by the redshift factor $z(\tau)$, which is 
\be 
\frac{\rho_{BH}(\tau)}{\rho_{rel}(\tau)}=0.02(\epsilon_{12}\times M_6)^{4/3}.
\ee 
Now, the temperature of the relativistic background just before the black hole decay was 
\be 
T_{cool}\equiv T_{rel}(\tau)=T_{eq}/z(\tau)=183.5\epsilon_{12}^{-1/3}M_6^{-11/6} \rm{MeV}.
\ee 
The temperature of the particles produced during the BH decay is equal to
\be 
T_{BH}=\frac{m_{pl}^2}{8\pi M}=0.48\times 10^7M_6^{-1} \rm{GeV}.
\ee
Hence, the lightest beyond standard model particles with mass $m_X\sim 10^3$ GeV should be produced abundantly in process of PBH evaporation. 
\\
The average distance between PBH just before their decay was 
\be 
d^{BH}(\tau)=d^{BH}_{eq}.z(\tau)\approx 0.23\times 10^{-5} M_6^{7/3}\rm{cm}
\ee 
The total number of energetic particles produced by the decay of a single PBH is given by 
\be 
N_{hot}\approx \frac{M_{BH}}{3T_{BH}}=\frac{8\pi}{3}\Big(\frac{M}{m_{pl}}\Big)^2=1.8\times 10^{22}M_6^2.
\ee 
We assume the following model that the result of BH evaporation is a cloud of energetic particles with temperature as $T_{BH}$ and with radius $\tau_{BH}$ given by 
\be 
\tau_{BH}=4.5 M_{6}^3\rm{cm}.
\ee 
The number of PBHs in this common cloud is 
\be 
N_{cloud}=(\tau_{BH}/d_{BH}(\tau))^3=7.593\times 10^{18}M_6^2
\ee 
and their number density just before the decay was 
\be 
n_{BH}(\tau)=d(\tau)^{-3}=8.21\times 10^{16}M_6^{-7}\rm{cm}^{-3}.
\ee 
The density of hot particles with temperature $T_{BH}$, created by evaporation of this set of black holes is 
\be
\label{nhot}
n_{hot}=n_{BH}.N_{hot}=14.78\times 10^{38}M_6^{-5}\rm{cm}^{-3}.
\ee 
The density of cool background particles with temperature $T_{cool}$ is 
\be 
0.1g*T_{cool}^3=3.03\times 10^{37} \epsilon_{12}^{-1}M_8^{-11/2} \rm{cm}^{-3}
\ee 
where we took $g_*=10$ at $T<100$\rm{MeV}.\\
The cooling proceeds through the Coulomb-like scattering and hence the momentum of hot particles decreases according to the equation 
\be 
\dot{E}_{hot}=-\sigma vn_{cool}\delta E,
\ee 
where $\delta E$ is the momentum transfer from hot particles to the cold ones. For massless particles, 
\be 
q^2=(p_1-p_2)^2=-2(E_1E_2-p_1.p_2). 
\ee
Finally we have 
\be 
\dot{E}=0.1g*T^3_{cool}\alpha^2/E_1\approx 10^{-4}T_{cool}^2.
\ee 
The loss of energy of hot particles of the order of their temperature would be achieved during very short time 
\be 
t_{cool}\approx 10^{-10} \textrm{sec}
\ee 
As a result of mixing and thermalization between the hot and cool components, the temperature of the resulting plasma would be 
\be 
\label{final}
T_{fin}=T_{cool}(\rho_{hot}/\rho_{cool})^{1/4}\approx 69.007 M_6^{-3/2} \rm{MeV}.
\ee 
Then, the total number density of relativistic particles would be equal to 
\be 
n_{rel}=0.1 g*T^3_{fin}=328608.99 M_6^{-9/2} (MeV)^3
=0.04\times 10^{39}M_6^{-9/2} cm^{-3}
\ee 
According to (\ref{nhot}), the number density of $X$-particles immediately after evaporation should be about $10^{39}M_6^{-5} cm^{-3}$. After fast thermalization, the ratio of number densities of $X$s to that of all relativistic particles becomes
\be 
n_X/n_{rel}=35
\ee 
The evolution of the number density of $X$-particles is given by the following equation 
\be 
\label{evol}
\dot{n}_X+3Hn_X=-\sigma^{ann}_Xvn^2_X
\ee 
The Hubble parameter which enters (\ref{evol}) is given by the expression 
\be 
H=\Big(\frac{8\pi^3g*}{90}\Big)^{1/2}\frac{T^2}{m_{pl}}\approx \frac{0.4 T_{in}^2}{z^2m_{pl}}
\ee 
where $z=a_{in}/a$ is the ratio of the initial scale factor to the running one and for $T_{in}$, we take $T_{fin}$ given by (\ref{final}). Introducing $r=n_Xz^3$ and the changing the time variable to z, we arrive at 
\be 
\frac{dr}{dz}=-\frac{\sigma_{ann}v m_{pl}}{0.4 T^2_{in}}\frac{r^2}{z^2}
\ee 
which can be solved and we get 
\be 
n_X=\frac{n_{in}}{z^3(1-1/z)}\rightarrow \frac{1}{Qz^3},
\ee 
where $Q=(\sigma vm_{pl})/(0.4 T_{in}^2)$.
\\~\\
According to observational data, we have 
\be 
\Omega_{DM}=0.26~~ \textrm{and}~~ \Omega_{CMB}=5.5\times 10^{-5}
\ee 
or we can say in terms of energy density that $(\rho_X/\rho_{\gamma})_{obs}\approx 5\times 10^3$.
We now have 
\be 
\sigma_{ann}v m_{pl}\sim 3\times 10^{11}\textrm{GeV}^{-1}
\ee 
and 
\be 
n_X\approx 10^{-12}z^{-3}T_{in}^2 GeV
\ee 
Now, the red-shift factor $z$ depends on both $M_6$ and the initial energy density ($\epsilon$). Thus, an appropriate choice of both these parameters would yield a significant amount of dark matter relic. Indeed, if we choose $\epsilon = 10^{-14}$, the resultant number density of $X$ (DM particles) is given by 
\be 
n_X \approx 1.1\times 10^3~cm^{-3}
\ee 
In the following tables we have shown the variation of the ratio of the dark matter particles to the relativistic particles for different values of the ratio of the mass of PBH to that of the mean mass of the spectrum. For brevity, the notation is kept same as $M_6$. The parameter space for the Extended mass spectrum-II is kept same as that of the first one in order to lay out the comparison more clearly. 

\begin{center} 
\begin{tabular}{ |p{3cm}|p{3cm}|p{3cm}|  }
\hline
\multicolumn{3}{|c|}{Extended mass spectrum-I} \\
\hline
$M_6$& $n_{BH}(\rm{cm}^{-3}$)& $n_X/n_{rel}$ \\
\hline
$0.1$ & $8.21 \times 10^{23}$ & 110.68  \\ 

 $1$ & $8.21 \times 10^{16}$ & 35  \\ 

 $10$ & $8.21 \times 10^{9}$ & 11.068 \\
 $100$ & $8.21 \times 10^{2}$ & 3.5 \\
\hline
\end{tabular}
\end{center}

\begin{center}
\begin{tabular}{ |p{3cm}|p{3cm}|p{3cm}|  }
\hline
\multicolumn{3}{|c|}{Extended mass spectrum-II} \\
\hline
 $M_6$ & $n_{BH}(\rm{cm}^{-3})$ & $n_X/n_{rel}$ \\ [0.5ex] 
 \hline
 $0.5$ & $1.05 \times 10^{19}$ & 49.49  \\ 
 
 $1$ & $8.21 \times 10^{16}$ & 35  \\ 
 
 $20$ & $6.41 \times 10^{7}$ & 7.826 \\
 
 $150$ & $0.48 \times 10^{2}$ & 2.857 \\
 \hline
\end{tabular}
\end{center}

As can be seen from the two tables the ratio of the relic abundance to that of the relativistic particles is different for both spectra, becaause their parameter spaces are chosen to be different. However, it is not hard to see that the ratio only depends on the peak value of the PBH spectrum and not on the entire distribution.

\begin{figure}[!htbp]
  \centering
  \begin{minipage}[b]{0.45\textwidth}
    \includegraphics[width=\textwidth]{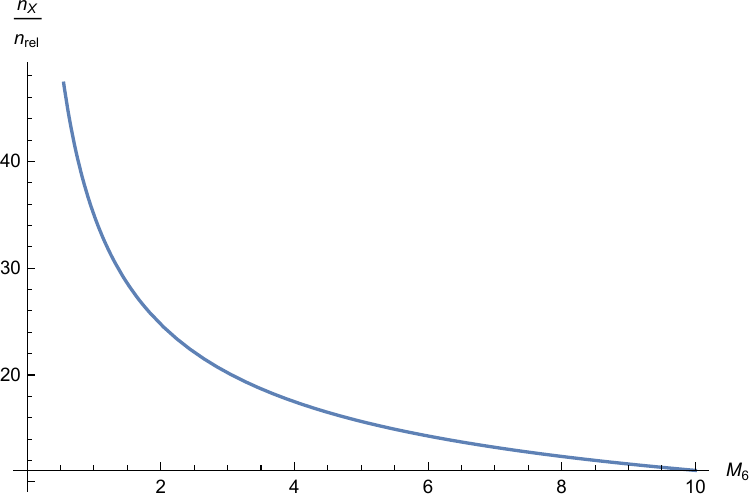}
  \end{minipage}
  \hspace*{.1cm}
  \begin{minipage}[b]{0.45\textwidth}
    \includegraphics[width=\textwidth]{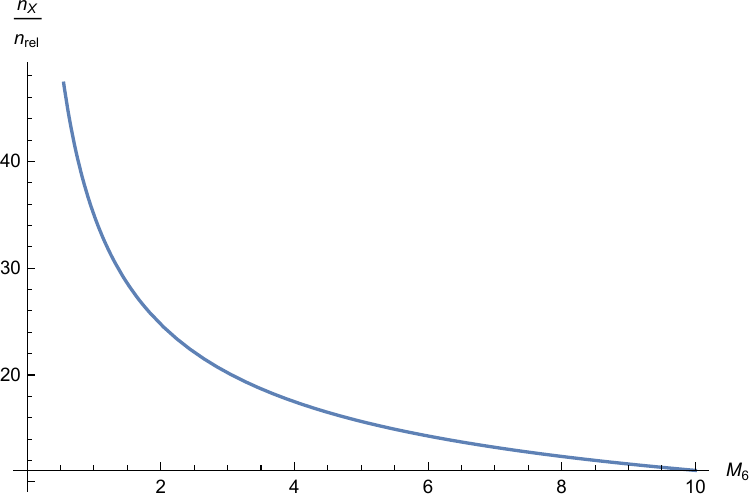}
      \end{minipage}
    \hspace*{.1cm}
  \caption{The variation of the ratio of the relic to that of the relativistic particles produced as a result of PBH evaporation is shown. The left panel corresponds to that of Extended Mass Spectrum-I and the right panel to that of Extended Mass Spectrum-II.
 } \label{ratio}
 \end{figure}
As can be seen from the above figure (\ref{ratio}) the two panels are identical and hence showing the production of relics due to PBH evaporation is not dependent on the mechanism of the production of PBHs. It is also clear that the the ratio of the mean mass of the black hole to the peak value of the black hole is inversely proportional to the relic abundance. Lesser the ratio, greater is the relic abundance.

\begin{figure}[!htbp]
  \centering
  \begin{minipage}[b]{0.45\textwidth}
    \includegraphics[width=\textwidth]{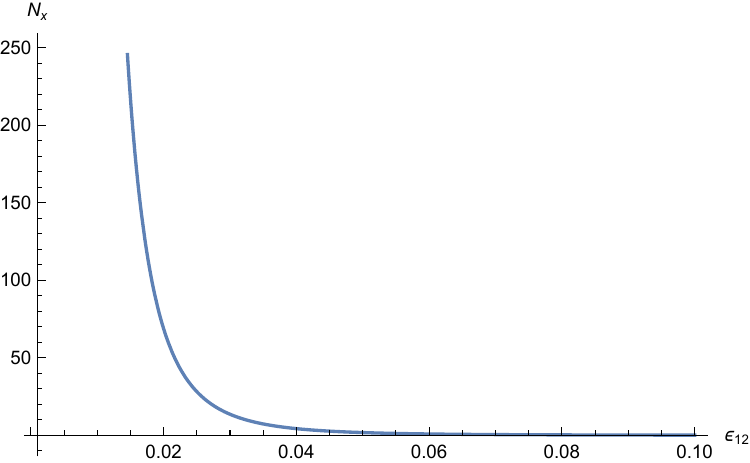}
  \end{minipage}
  \hspace*{.1cm}
  \begin{minipage}[b]{0.45\textwidth}
    \includegraphics[width=\textwidth]{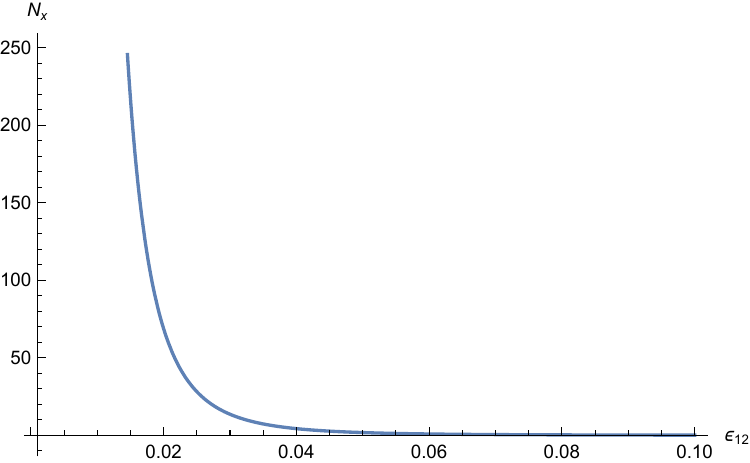}
      \end{minipage}
    \hspace*{.1cm}
  \caption{The production of dark matter particles as a result of PBH evaporation is shown. The left panel corresponds to that of Extended Mass Spectrum-I and the right panel to that of Extended Mass Spectrum-II.
 } \label{DM}
 \end{figure}
It is clear from the above figure (\ref{DM}) the relic abundance depends on the energy density of the primordial black holes. The dependence is given by the following relation:
\be 
n_X \sim \frac{1}{\epsilon_{12}^{4}}
\ee 
Thus, lesser the initial energy density of the primordial black holes, more is the relic abundance. Since the ratio of the DM particles to that of relativistic matter is again proportional to the inverse fourth power of the initial energy density, we can get a significant ratio $(>1)$ by adjusting the value of the initial energy density appropriately.

\section{Conclusion}
Production of primordial black holes can lead to a specific form of their mass spectrum, reflecting physical mechanism of their creation. Such mechanisms relate the mass interval and specific features of the PBH spectrum to the BSM physics at very high energy scales. The results of our work highly depends on the values of the parameters in (\ref{LN}). Even though the nature of the distribution won't be affected by changing the values of the parameter but the value of $x$ corresponding to the peak value or the maximal value of the distribution will change if the parameters are modified. Thus, this would not modify our result as it depends on the peak value of the distribution. In other words, changing the values of $a$ and $b$ will skew the distribution towards the left or the right. \\~\\
In the recent work \cite{kousik}, it is shown how the evaporation of a singular PBH, which can be approximated as a delta function mass spectrum, can lead to the instant change of regime and contribute to the dark matter density of the universe. It was shown how the energy density of the PBH affected the density of the dark matter. On the same note, it is established in this work that the initial energy density of the PBHs at the very early stage of the universe expansion plays a significant role in the production of dark matter (relic abundance) as a result of PBH evaporation. This idea has been tested for the cases when PBHs were formed by the ZN mechanism accomplished by strongly increased probability of their formation in some interval of small PBH mass and also for the case when they were formed by the collapse of domain wall. In both the cases, the resulting number density of the dark matter particles is significant. We conclude with the claim that the dark matter production process due of the evaporation of PBHs does not depend of the mechanism of PBH formation for the same peak mass in their mass distribution.  However, the PBH mass spectrum may be flat as it is the case for PBH formation at the post-inflational matter dominant stage of massive scalar field \cite{khlopov1}. There is no evident peak mass value in such case and the proper analysis of evolution of PBH dominant stage and DM production in PBH evaporation need special study, which we plan in our future work.

\section{Acknowledgments}
The authors are thankful to Kousik Loho and Baradhwaj Coleppa for useful discussions. Further, the authors express their gratitude to A.S. Sakharov, J. Turner and her group and Vincent Vennin for pointing towards some recent and interesting work on the production of relic particles and gravitational waves as a result of PBH evaporation. The work of A.C. is supported by the project RES/SERB/PH/P0202/2021/0039. The work by M.K. was performed with the financial support provided by the Russian Ministry of Science and Higher Education, project “Fundamental and applied research of cosmic rays”, No.~FSWU-2023-0068. The work of P.C. is supported by the EPSRC fellowship.

\end{document}